\documentclass[epsf,usegraphicx]{mn2e}

\usepackage{xspace}
\usepackage{amsfonts}
\usepackage{pifont}

\title[Questioning the traditional view of early dM stars]{The view
  from K2: Questioning the traditional view of flaring on early dM
  stars}

\author[]
{Gavin Ramsay$^{1}$, J. Gerry Doyle$^{1}$ \\
$^{1}$Armagh Observatory, College Hill, Armagh, BT61 9DG, UK\\
}

\date{Accepted 2015 March 9.  Received 2015 March 2; in original form 2014 October 27}

\begin{document}
\outer\def\gtae {$\buildrel {\lower3pt\hbox{$>$}} \over 
{\lower2pt\hbox{$\sim$}} $}
\outer\def\ltae {$\buildrel {\lower3pt\hbox{$<$}} \over 
{\lower2pt\hbox{$\sim$}} $}
\newcommand{\ergscm} {ergs s$^{-1}$ cm$^{-2}$}
\newcommand{\ergss} {ergs s$^{-1}$}
\newcommand{\ergsd} {ergs s$^{-1}$ $d^{2}_{100}$}
\newcommand{\pcmsq} {cm$^{-2}$}
\newcommand{\ros} {\sl ROSAT}
\newcommand{\chan} {\sl Chandra}
\newcommand{\xmm} {\sl XMM-Newton}
\newcommand{\kep} {\sl Kepler}
\def\rchi{{${\chi}_{\nu}^{2}$}}
\newcommand{\Msun} {$M_{\odot}$}
\newcommand{\Mwd} {$M_{wd}$}
\newcommand{\Lsol} {$L_{\odot}$}
\def\Mdot{\hbox{$\dot M$}}
\def\mdot{\hbox{$\dot m$}}
\newcommand{\teff}{\ensuremath{T_{\mathrm{eff}}}\xspace}
\newcommand{\tickYes}{\checkmark}
\newcommand{\tickNo}{\hspace{1pt}\ding{55}}
\newcommand{\srcone} {TYC1330}
\newcommand{\srctwo} {RXJ0626}

\maketitle

\begin{abstract}
We use K2 short cadence data obtained over a duration of 50 days
  during Campaign 0 to observe two M1V dwarf stars, TYC 1330-879-1 and
  RXJ 0626+2349. We provide an overview of our data analysis, in
  particular, making a comparison between using a fixed set of pixels
  and an aperture which follows the position of the source. We find
  that this moving aperture approach can give fewer non-astrophysical
  features compared to a fixed aperture. Both sources shows flares as
  energetic as observed from several M4V stars using both {\kep} and
  ground based telescopes. We find that the flare energy distribution
  of the sources shown here are very similar to the less active M3--M5
  stars but are $\sim$8 times less likely to produce a flare of a
  comparable energy to the more active M0--M5 stars. We discuss the
  biases and sources of systematic errors when comparing the activity
  of stars derived from different instruments. We conclude that K2
  observations will provide an excellent opportunity to perform a
  census of flare activity across the full range of M dwarf spectral
  class and hence the physical mechanisms which power them.
\end{abstract}


\section{Introduction}

The {\kep} satellite provided a unique resource to study the
  activity levels of stars over a wide spectral type. Such
  observations can help address questions such as why there is a
  marked change in activity levels around spectral type M4 (West et
  al. 2008).  We initiated a campaign using {\kep} in Q14 to observe M
  dwarfs over a range of spectral type and in Ramsay et al. (2013) we
  reported observations of two M4V stars which showed strong flare
  activity.

However, the loss of two of {\sl Kepler's} four reaction wheels
limited the accuracy with which it could be pointed, with a resulting
degradation of the photometric accuracy. NASA, in collaboration with
the spacecraft manufacturers Ball Aerospace, were able to re-design
the mission so that by pointing at fields in the ecliptic plane, the
roll of the spacecraft is minimised. Pointing stability is achieved by
the firing of micro-thrusters. The {\kep} team found that K2 (which
the mission has now been renamed) can deliver photometry which is
accurate to within a factor of 2--4 of the original {\kep} data (see
Howell et al. 2014, Vanderburg \& Johnson 2014 and Aigrain et al
2015).

K2 will make a series of pointings each lasting $\sim$75 days along
the ecliptic plane -- some of the fields will be close to the Galactic
plane. In the planning stage for the K2 mission, several engineering
tests were made. Data from the Feb 2014 tests have been made
publically available and has already led to the publication of papers
on a number of objects, including a pulsating sdB star (Jeffery \&
Ramsay 2014), a pulsating white dwarf (Hermes et al. 2014) and a
sample of eclipsing binary stars (Conroy et al. 2014).

K2 will provide a unique set of observations which (with additional
complementary data) will provide a basis for determining how the
activity of stars, including late type dwarfs, is related to mass,
rotation, age and metalicity. During an engineering test of K2 early
in 2014, observations were made of the M4.5V flare star AF PSc with a
cadence of 30 min (Ramsay \& Doyle 2014). These tests showed step-like
features could be present in the light curves due to limitations in
the pointing stability, and that care had to be taken to correctly
distinguish stellar flares from instrumental artifacts.

Here we present K2 observations of two M1V stars in `Campaign 0',
which were the only M dwarf stars to be observed with 1 min cadence in
this campaign.  We outline the issues we faced in analysing the data
and make a comparison with light curves extracted using {\kep}
software and a well established photometric package where the aperture
was allowed to move over the pixel array during the course of the
observation.  We compare the flare activity of these M1V stars with
the previously observed M4V stars together with a range of M dwarfs of
varying activity levels as observed from ground based telescopes.

\begin{table*}
\begin{center}
\begin{tabular}{lcrrrrrrrr}
\hline
Star & EPIC & RA  & DEC  & Kep & Spectral & 
Spec Parallax & Phot Parallax & H$\alpha$ EW & Period\\
     &  ID    &  (J2000)          & (J2000)            & Mag & Type & (pc) & (pc) & (\AA) & (days)\\
\hline
TYC 1330-879-1 & 202059229 & 06:46:45.6 & +15:57:42.1 & 10.5 & M1.0V & 22.7$\pm$4.8 & 16.7$\pm$3.9 & -0.6 & 5.04\\
RXJ 0626+2349 & 202059204 & 06:26:14.5 & +23:49:28.5 & 11.3 & M1.5V & 28.6$\pm$5.9 & 31.3$\pm$6.9 & -0.7 & 7.90\\
\hline
\end{tabular}
\end{center}
\caption{We show the K2 Ecliptic Plane Input Catalog (EPIC) number of
  our targets; their sky co-ordinates; their KepMag (all taken from
  the EPIC catalog which can be accessed from
  http://archive.stsci.edu/k2/epic). The optical spectral type,
  spectroscopic and photometric parallaxes and the equivalent width of
  the H$\alpha$ line are all taken from L\'{e}pine et al. (2013). The
  rotation periods are derived from the K2 data.}
\label{sources}
\end{table*}

\section{Targets}

We performed a search for late type dwarfs in the Campaign 0 field of
view. The catalogue of bright M dwarfs of L\'{e}pine et al. (2013)
gave two sources which were in Field0 and `on-silicon'. TYC 1330-879-1
(hereafter {\srcone}) is a M1.0V dwarf ($V$=11.6) and 1RXJ
062614.2+234942 (hereafter {\srctwo}) is an M1.5V dwarf ($V$=11.8).
{\srcone} is $\sim$20 pc distant and {\srctwo} is $\sim$30 pc distant.
(see Table \ref{sources} for the key parameters for both sources).

\section{K2 Data}

The K2 Campaign 0 was carried out between MJD = 56728.0 --
  56804.7 (2014 Mar 12 -- 2014 May 27). However, there were two gaps
  in the data and the resulting coverage was 50.0 days. (This compares
  with 8.9 days for the engineering test data). We call the three
  contiguous sets of data sections 1, 2 and 3.  During sections 1 and
  2, K2 was operated in `coarse' pointing, whilst during section 3 it
  was operated in `fine' pointing during which the resulting
  photometric precision being a factor of two
  better\footnote{http://keplerscience.arc.nasa.gov/K2/Performance.shtml}.
  Observations of {\srcone} and {\srctwo} were made in {\it short
    cadence} (SC) where the effective exposure is 58.8 sec ({\it long
    cadence data} which has an exposure of 30 min is also available).

During Campaign 0 a 50$\times$50 pixel array was downloaded for each
target and 73470 individual images were obtained for each source.
To correct for the drift in the satellite pointing, thrusters are
  used to periodically re-saturate the reactions wheels. This results
  in a significant movement in the targets position on the array every
  $\sim$2 days and it can take around a dozen or more SC images for
  the pointing to stabilise.  Vanderburg \& Johnson (2014) outline a
technique to remove the effects of correlated systematic
variations. As a service to the community they have provided light
curves for all LC data (but not SC data) taken during Campaign 0.

Since data reduction of K2 data is more complex than that of {\kep}
data, we have compared the results of extracting light curves from the
SC datasets using the {\tt PyKe} software (Still \& Barclay
2012)\footnote{http://keplergo.arc.nasa.gov/PyKE.shtml} which was
developed for the {\sl Kepler} and K2 mission by the Guest Observer
Office with a well established aperture photometric package ({\tt
  AUTOPHOTOM}, Eaton, Draper \& Allan 2009) which is part of the
STARLINK software
collection\footnote{http://starlink.jach.hawaii.edu/starlink}.  In
each approach we analysed each section separately then combined the
resulting light curves. We used K2 data from Data Release 2.

\begin{figure*}
\begin{center}
\setlength{\unitlength}{1cm}
\begin{picture}(8,10)
\put(13.6,-0.6){\includegraphics{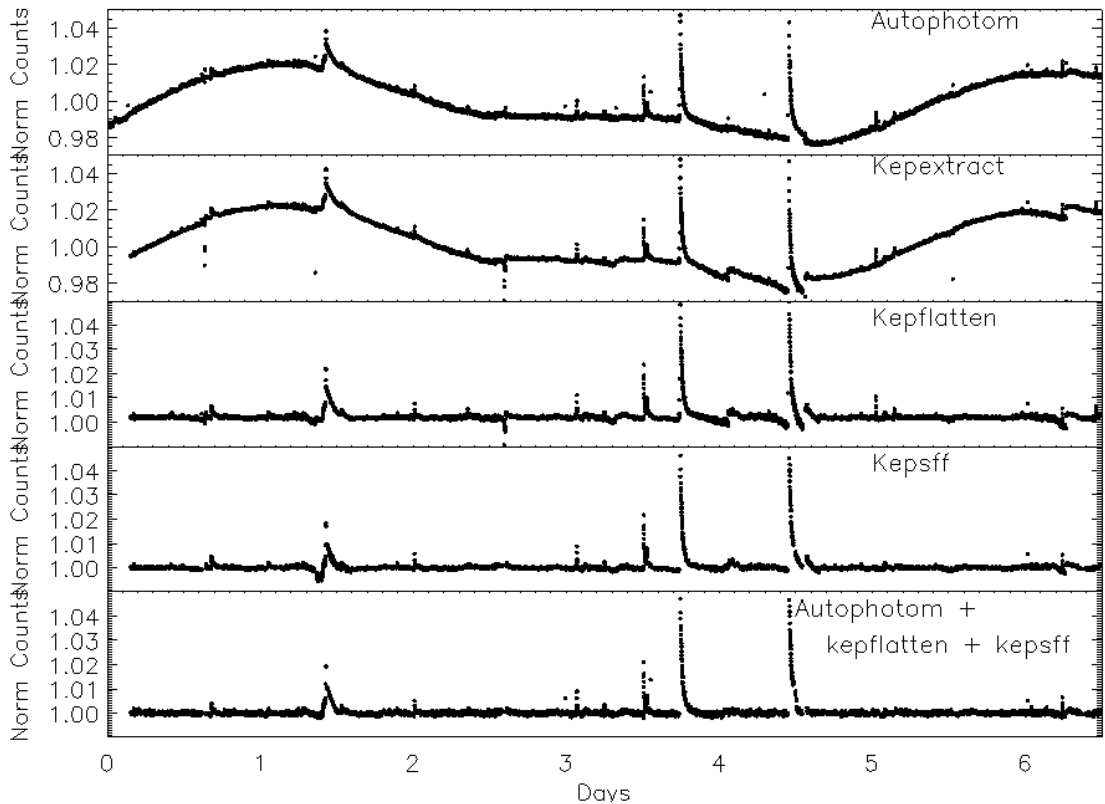}}
\end{picture}
\end{center}
\caption{The light curve of {\srcone} derived from section 2 data made
    using different approaches and stages in the analysis. In the
    upper panel we show the light curve made using a moving fixed
    aperture photometric package ({\tt Autophotom}). In the lower
    panels we show the light curve derived using the {\tt PyKe} tool
    {\tt kepextract} followed by the removal of the long term trend
    ({\tt kepflatten}) and the removal of systematic trends using {\tt
      kepsff}. In the bottom panel we show the light curve derived
    using {\tt Autophotom} after long term trend and systematic trends
    have been removed.}
\label{section2}
\end{figure*}

Using {\tt kepmask} and {\tt kepextract} (which are part of the {\tt
  PyKe} suite of tools) we extracted a light curve using data from
pixels centered on the target. We explored using different numbers of
pixels and found that clear discontinuities are seen when smaller
number of pixels are used. However, beyond a certain number of pixels,
the noise in the resulting light curve increases. At this stage in the
reduction, a distinctive modulation in each target star was apparent,
which we took to be the stellar rotation period. To remove the effects
of this modulation we used {\tt kepflatten}. It is now possible to
remove much of the correlated noise in the light curves using {\tt
  kepsff} which was developed using the method outlined in Vanderburg
\& Johnson (2014).  The light curve produced after each step of the
process is shown in Figure \ref{section2}. Data have been normalised
so that it is divided by the mean of each section of data.
Considerable time was spent in selecting optimal parameters for the
different {\tt PyKe} tasks. However, as is clear from the second lower
panel of Figure \ref{section2}, there are features in the light curve
which may due to the result of an imperfect removal of the signature
of the stellar rotation period or the presence of residual systematic
trends in the data.

We then used {\tt AUTOPHOTOM} which has been used by several groups
over many years to analyse ground based images (Eaton, Draper \& Allan
2009). Images were extracted from the SC data file using {\tt
  kepimages}. Epochs of thruster events were easily identified since
the mean pixel value for these events was zero. These images and the
next 24 images were not considered further in our analysis since the
spacecraft pointing took some time to stabilse.

To extract photometry we used an aperture of fixed size and allowed
the center of the aperture to track any movement of the star across
the pixel array. We also defined a source free region which tracked
the movement of the source aperture to subtract the background
flux. We explored using an aperture of different radii and found that
an aperture of 6 pixel radius gave the best compromise between having
fewer discontinuties in the data and low rms of the light curve.  A
small correction is then applied to each shorter sub-section of data
so that no discontinuties are present between these sub-sections.  We
show the light curve derived from section 2 data using {\tt
  AUOTPHOTOM} in the top panel of Figure \ref{section2}. Compared to
the light curve derived using {\tt kepextract} we find fewer
discontinuties. We also removed the long term trend due to stellar
rotation and also the effects of correlated noise using the {\tt PyKE}
task {\tt kepsff} as done before. Although the rms of the resulting
light curves derived using {\tt AUTOPHOTOM} and {\tt PyKE} tasks are
very similar, the former has fewer features which are likley to be
instrumental. For instance, the last 2 days of the light curve
  shown in Figure \ref{section2} (i.e. immediately after the flare at
  Day $\sim$4.5 to the end) has an rms of 0.084 percent using {\tt
    AUTOPHOTOM} (the bottom panel of Figure \ref{section2}) compared
  to 0.089 percent using {\tt PyKE} (second from bottom panel).  For
the remainder of the analysis we used the light curves derived using
{\tt AUTOPHOTOM} and subsequent removal of the long term trends due to
stellar rotation.

\begin{figure*}
\begin{center}
\setlength{\unitlength}{1cm}
\begin{picture}(8,13)
\put(15,-0.8){\includegraphics{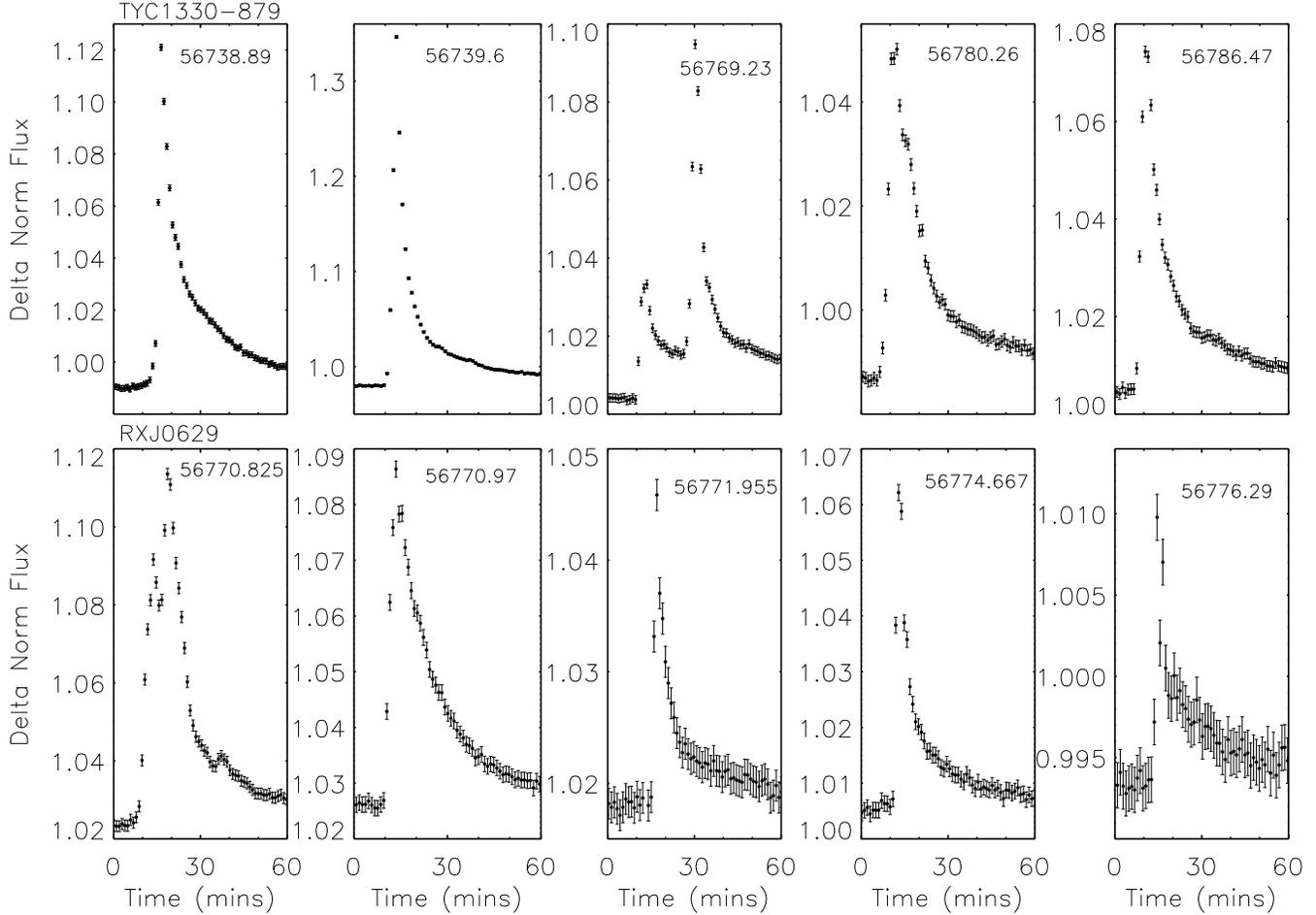}}
\end{picture}
\end{center}
\caption{The five most energetic flares seen in K2 SC data of
  {\srcone} (top panels) and {\srctwo} (lower panels). They have been
  arranged in chronological order.}
\label{flares}
\end{figure*}

In previous papers we have pin-pointed flares in the light curve by
flagging all times which were above a certain threshold above the
mean. However, given the presence of discontinuities in some parts
of the light curves, in this study we identified potential flares by
examining the light curves by eye. This is a tractable approach since
our sources do not exhibit vast numbers of flares and we have only two
sources to examine. Moreover, since we had two simultaneous SC light
curves, if an event was seen in both light curves it was deemed
instrumental and not considered further, as was also the case for
events consisting of one time point.  As {\tt kepsff} identifies and
flags times of thruster events (and other instrumental effects), we
removed any event which was within two dozen images of these events
using the {\tt SAP\_QUALITY} flag. For both stars, 31 flares were
detected. We show the five most luminuous flares identified in each
star in Figure \ref{flares}.

Each of the events which we have identified as flares have a
characteristic stellar flare profile -- i.e. a sharp rapid rise to
maximum followed by a quasi-exponential decay. The rise to maximum has
been resolved and typically takes 1--5 mins. The duration of the
flares are typically $\sim$10--20 mins, while in one event, ({\srcone}
MJD=56769.23) two flares are seen in rapid succession, while in
{\srctwo} (MJD=56770.825) the profile of the peak in more complex than
others (this event was recorded as one flare). It is clear that none
of these flares would have been resolved in long cadence mode data.

\section{Results}
\label{results}

\subsection{Rotation Period}

For both light curves we removed any trends which were longer
  than the stellar rotation period. Given that various offsets have
  been applied to sub-sections of data more detailed work would be
  required to search for evidence of differential rotation in these
  light curves (see Lurie et al (2015) for evidence for such an effect
  in the M5V star GJ 1245B). We determined the rotation period of each
star using the standard Lomb Scargle periodogram. We found a rotation
period of 5.04 days for {\srcone} and 7.90 days for {\srctwo}. We
show in the top panels of Figure \ref{energy-phase} the K2 data folded
on the rotation period where phase 0 has been chosen to correspond to
minimum flux.

{\srcone} show a peak-to-peak amplitude of 2.7 percent and 2.3 percent
for {\srctwo}. Both light curves show a minimum which is likely
due to an enhancement in the number of spots which are darker than the
surrounding photosphere and are visible at this rotational
phase. There is also a local minimum in the light curve of both stars
but is much deeper in {\srctwo}. This may suggest that the star spots
have two distinct locations (perhaps one in each hemisphere).

Nielsen et al. (2013) and McQuillan et al. (2013) determine the
rotation period of main sequence stars using {\kep} data. They find
that there is a general relationship between mass and rotation period
such that the rotation period increases towards lower masses. However,
for any given spectral class there is a large spread in rotation
period: Kiraga \& Stepien (2007) show that for stars with a mass
between 0.5--0.6 \Msun (stars with a spectral type M1V has a mass
$\sim$0.56 \Msun, Baraffe \& Chabrier (1996)) the rotation period can
be a fraction of a day to tens of days.  It is likely that other
factors (such as age) effect the rotation period.

\subsection{Flare Energies}

We identified a total of 31 flares in both {\srcone} and {\srctwo}.
To determine the luminosity of these flares we estimated the
luminosity of the two sources using the relationship from L\'{e}pine
\& Gaidos (2011):

\begin{equation}
M_{V}\sim 2.2 (V-J) + 2.5
\end{equation}

where (V-J) = 3.66 and 3.14 (also taken from L\'{e}pine \& Gaidos) and
implies $M_{V}$=10.55 and 9.4 for {\srcone} and {\srctwo}
respectively. We assume the Sun has $M_{V}$=4.83 and
$L=3.8\times10^{33}$ erg/s which implies $L=2.0\times10^{31}$ erg/s
and $L=5.6\times10^{31}$ erg/s for {\srcone} and {\srctwo}
respectively. For each flare we then measured the amount of energy per
time bin by comparing the flare energy with the quiescent flux level
and the energy of the flare was summed up. The range of flare
  energy in {\srcone} was 2$\times10^{31}-6.6\times10^{33}$ erg and in
  {\srctwo} 7$\times10^{31}-5.1\times10^{33}$ erg. We estimate the
spread in the (V-J), $M_{V}$ relationship of L\'{e}pine \& Gaidos
(2011) is $M_{V}\sim$0.5 which translates to an uncertainty on the
resulting luminosity of $\sim$ 40 percent.

\begin{figure}
\begin{center}
\setlength{\unitlength}{1cm}
\begin{picture}(8,6)
\put(9.4,-0.6){\includegraphics{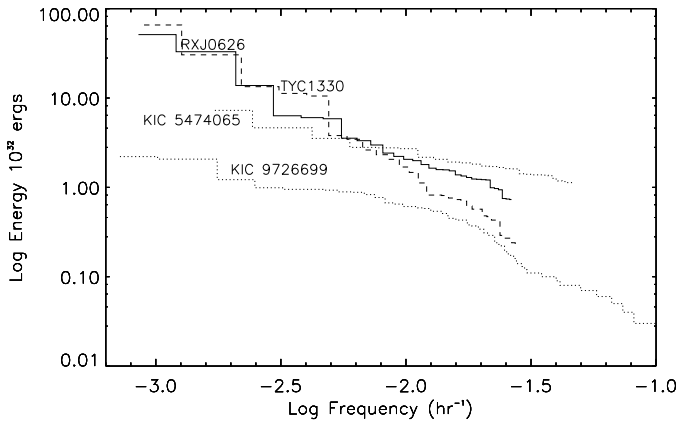}}
\end{picture}
\end{center}
\caption{The cumulative energy distribution of flares (in the {\kep}
  band-pass) as seen in {\srcone} and {\srctwo}. We also show the
  distribution of flare energy of KIC 5474065 and KIC 9726699 using
  {\kep} data (Ramsay et al. 2013).}
\label{cumulative}
\end{figure}

In Figure \ref{cumulative} we show the cumulative energy distribution
of flares seen in {\srcone} and {\srctwo} together with those seen in
KIC 5474065 and KIC 9726699 (Ramsay et al. 2013), which indicates how
often a flare with a given energy is seen. Both {\srcone} and
{\srctwo} show flares with energies $L\sim10^{33}$ erg roughly every 8
days which is roughly twice as frequent as KIC 5474065.  On the other
hand {\srcone}, {\srctwo} and KIC 5474065 show the same frequency of
flares with energies a few $\times10^{32}$ ergs. Given the uncertainty
in the luminosities, the slope of flare energy distribution per time
of each K2 source is similar.

\begin{figure}
\begin{center}
\setlength{\unitlength}{1cm}
\begin{picture}(8,7)
\put(9.4,-0.6){\includegraphics{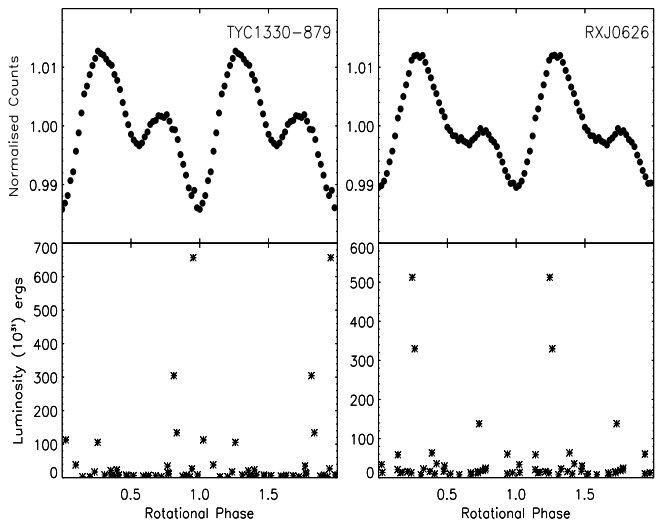}}
\end{picture}
\end{center}
\caption{The light curve of {\srcone} (left hand panels) and
    {\srctwo} (right hand panels) folded on the ephemeris $T_{o} (MJD)
    =$56724.74(6)+5.036(5) and $T_{o} (MJD) =$56721.5(1)+7.90(2)
    respectively where the number in parenthesis is the error on the
    last digit and $\phi$=0.0 has been defined as minimum flux. (The
    data has been plotted over two rotational cycles for clarity). In
    the lower panels we show the energy of the flares as a function of
    rotational phase.}
\label{energy-phase}
\end{figure}

In Figure \ref{energy-phase} we show the light curve of both sources
folded on the stellar rotation period. We also show the energy of the
flares as a function of rotational phase where we define $\phi$=0.0 as
the point of minimum flux. This indicates that while flares are
  seen at all rotational phases, the most energetic flare seen in
  {\srcone} is close to minimum flux. This is what we would expect if
the flares originate from active regions close to starspots, which
being cooler give a lower flux when they are visible. On the
  other hand the most energetic flare seen in {\srctwo} is seen close
  to maximum brightness.

\section{Discussion}

The K2 mission will allow the determination of rotation period and
flare rate for a wide range of M type dwarf stars and assess how these
factors are related with mass and age. We find that for the two
sources with very similar spectral type (M1V and M1.5V), the flare
characteristics are similar, with more energetic flares compared to
the two M4V stars (Figure \ref{cumulative}). This is intriguing since
the activity levels of M dwarfs (as measured for instance by levels of
H$\alpha$ emission) has been found to be very low from M0V stars,
reaching 40 percent of M5V stars being active while 90 percent of
stars later than M5V are found to be active (e.g. West et al. 2008 and
Schmidt et al. 2014).

In Ramsay et al. (2013) we compared the activity levels of the two M4V
dwarfs KIC 5474065 and KIC 9726699, as measured using {\kep} data,
with other stars. To do this we estimated the equivalent energy of the
flares in the U band (where many observations have been made). As
before, we assume $E_{Kepler}/E_{U}=2.4$. We therefore find that
  for {\srcone}, flares have $L_{U}\sim1\times10^{31}-3\times10^{33}$
  erg and for {\srctwo} $L_{U}\sim3\times10^{31}-2\times10^{33}$ erg.

\begin{figure*}
\begin{center}
\setlength{\unitlength}{1cm}
\begin{picture}(8,11.8)
\put(-4.5,12.5){\includegraphics{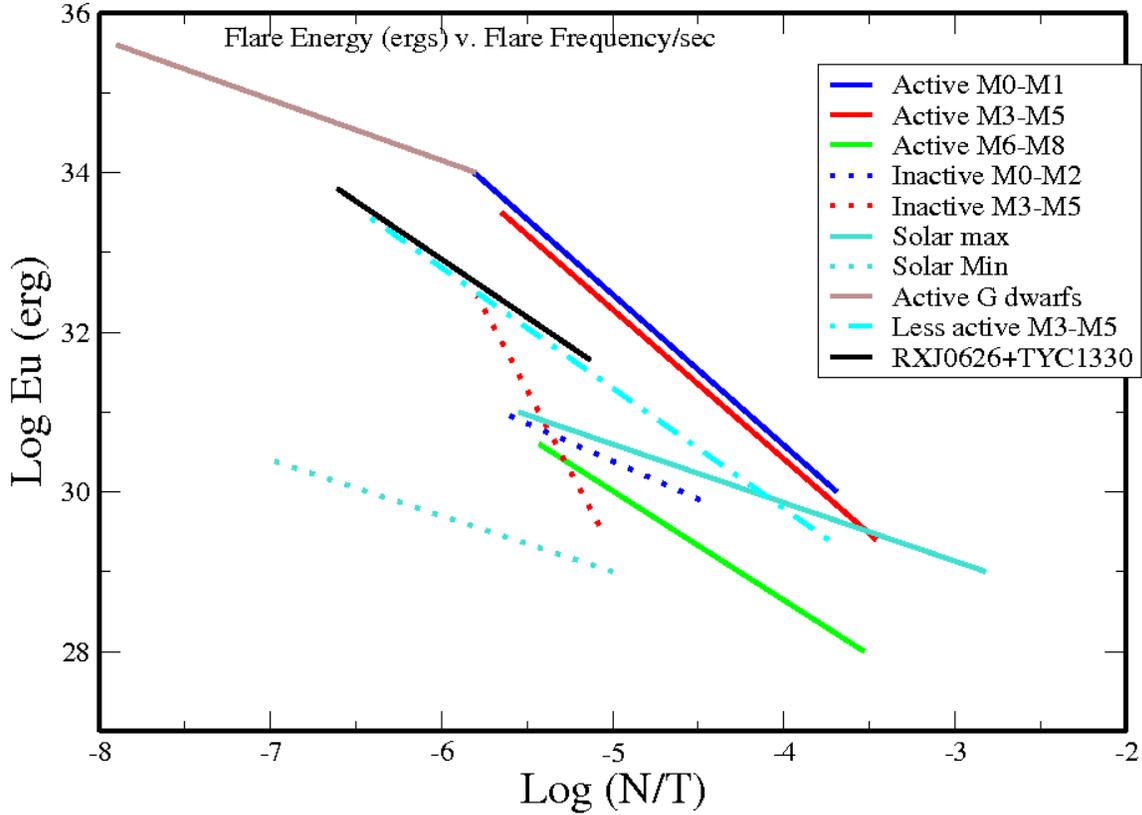}}
\end{picture}
\end{center}
\caption{The cumulative flare frequency (in seconds) versus U-band
  flare energy (in erg) for different classes of dwarf stars and
  {\srcone} and {\srctwo}. Data is taken from Hilton (2011): Active
  M3-M5, 332 hours on 4 stars observing 157 flares; Active M6-M8, 59
  hours on 4 stars observing 39 flares; Less Active M3--M5, 147 hrs on
  8 stars observing 28 flares; Inactive M0-M2, 256 hours on 16 stars
  observing 9 flares; Inactive M3-M5, 153 hours on 6 stars observing 3
  flares; Active M0-M1, data from Moffett (1974), Doyle \&
  Mathiouidakis (1990) and Dal \& Evren (2011), 156 hours on 2 stars
  observing 63 flares. In addition, we include from Shibayama (2013) a
  line for solar maximum and minimum which used a bolometric/GOES
  X-ray flux relation derived from work by Kretzschmar (2011). Data
  for active G dwarfs are taken from Shibayama et al (2013) which took
  46,000 hrs of observations on 4 stars observing 116 flares.}
\label{Uband-rates}
\end{figure*}

We produced a cumulative flare frequency based on plots such as those
shown in Fig \ref{cumulative} in Ramsay et al. (2013) where we derive
a linear relation of the form:\\
\begin{equation}
log (N/T) = a + b. log E
\end{equation}
We take the work of Hilton (2011) who determined the cumulative flare
frequency of active M3--M5 and M6--M8 stars together with inactive
M0--M2 and M3--M5 stars and less active M3-M5 (see Hilton (2011)
  and Hawley et al. 2014)). We supplement this by taking previous work
on active M0--M1 stars and also the Solar flare distribution for Solar
maximum and minimum (Figure \ref{Uband-rates} and see its caption for
references). The Sun at Solar minimum lies below the M dwarfs while at
Solar maximum it is consistent with several M dwarfs flare rate and
energy output.  Also included is data for active G dwarfs taken
  from Shibayama et al (2013).

The rate of flaring for events in TYC1330 and RXJ0626 has a
  similar slope as the active M0--M1 and active M3--M5 stars, but is
  $\sim$8 times less likely to produce a 10$^{33}$ erg flare. It is
  however in excellent agreement with the group of less active M3--M5
  dwarfs from Hilton (2011) and Hawley et al. (2014). The inactive
  M0--M2 stars from Hilton (2011) are much less active and do not
  produce flares above $10^{31}$ ergs.

In making a comparison between the cumulative flare rates of stars
derived from different instruments (Figure \ref{Uband-rates}) there
are a number of factors which should be noted. The first is the flare
energy calibration which as mentioned in \S 4.2 has a possible error
of around 40 percent. Another is the conversion from the Kepler filter
to $U$ band: we used $E_{Kepler}/E_{U}=2.4$, compared to
$E_{Kepler}/E_{U}=1.54$ used by Hawley et al. (2014) and
$E_{Kepler}/E_{U}=2.5$ implied from observations by Hawley \&
Pettersen (1991). Another is the accuracy of the power-law which can
result from a low flare count, in particular for the inactive
dwarfs. This is less of an issue with the present K2 data which has 62
flares from 2400 hours of observations. Although the slope for the
present K2 data is in excellent agreement with that for the less
active M3--M5 dwarf ground-based data, Hawley et al. (2014) noted that
using {\kep} data, the active M3--M5 dwarfs have a slope which is
steeper compared to ground based data.

Another uncertainty is the question of whether the flare star is
located in a binary system. For instance Rappaport et al (2014) find
that 17 percent of M dwarfs with rotation periods shorter than two
days show evidence for being in a binary or multiple system. At this
stage we do not know whether the K2 M1 stars (or indeed the {\kep} M4
stars) are in a binary system, but if they were this could give
enhanced activity.  However, perhaps the biggest unknown is the
question of activity cycles for these dwarfs: we note that if M dwarfs
show stellar magnetic cycles then their flare distributions would be
expected to move in the vertical axis of Figure \ref{Uband-rates} in
the same manner as the Sun. Work based on dM stars observed by {\kep}
coupled with new K2 data could help resolve this problem.

The two objects in the active M0--M1 branch are YY Gem (an M1/M1
binary) and V1005 Ori (M0V). YY Gem has a rotational period of 20
hours while there is no known rotational period for V1054
Oph. TYC1330-879 has a rotational period of 5 days. We speculate that
the lower flare activity rate of both TYC1330-879 and RXJ0626 is age
related with the slower rotators being older and less active. A
similar conclusion was made by West et al. (2008) based on MEarth
observations. Both YY Gem ($\sim$200 Myr) and V1005 Ori ($\sim$30 Myr)
are relatively young objects.

Optical spectroscopy shows that both {\srcone} and {\srctwo} have
  weak H$\alpha$ emission (equivalent width $<$1\AA, c.f. Table 1).
  Traditionally, M dwarfs with H$\alpha$ having an equivalent width of
  $<$1 \AA \hspace{1mm} were termed inactive, e.g.  West et
  al. (2008). The fact that these two objects produce flares may not
  be a surprise. However, the fact that they produce flares with an
  energy more than two orders of magnitude above the previous assumed
  limit for inactive M3--M5 stars is surprising. This shows the value
  of having an Earth orbiting instrument capable of long duration
  observations (i.e. months), compared with an Earth-based telescope
  and shows the potential impact that future K2 observations can make
  in the field of stellar flare observations.

\section{Conclusions}

We have compared the K2 short cadence light curves of two M1 V
  stars derived using a well established photometric package using a
  moving aperture with that derived using tools specifically written
  for {\kep} and K2 data. We find evidence for the moving aperture
  approach can give smoother light curves making it particularly
  suitable for identifying flares. Our targets show flares roughly
  once every two days, despite the fact that their low H$\alpha$
  emission would have classed them as inactive.  We compare their
  flare energy distribution with other M dwarfs observed using Kepler
  and find that TYC1330 and RXJ0626 shows more flares with energies of
  10$^{33}$ erg compared with our two comparison M4V stars as observed
  using Kepler in SC mode. Comparing their equivalent U band
  luminosity with other dwarfs, both TYC1330 and RXJ0626 show
  cumulative energy distributions with a similar slope as active
  M0--M5 stars, but their flaring rate is a factor of 8 lower. The K2
  mission will allow a wide range of late type dwarf stars to be
  targeted and assess their activity rates as a function of age and
  rotation period.

\section{Acknowledgements}

We would like to thank the anonymous referee for a constructive and
very useful report. Funding for the K2 spacecraft is provided by the
NASA Science Mission Directorate. The data presented in this paper
were obtained from the Mikulski Archive for Space Telescopes (MAST).
This work made use of PyKE, a software package for the reduction and
analysis of Kepler and K2 data. This open source software project is
developed and distributed by the NASA Kepler Guest Observer
Office. Armagh Observatory is supported by the Northern Ireland
Government through the Dept Culture, Arts and Leisure.

\end{document}